\documentclass[12 pt, onecolumn,showpacs,preprintnumbers,amsmath,amssymb]{revtex4}
\usepackage{graphicx}
\usepackage{dcolumn}
\begin{document}
\title{Stability of the Breached Pair State for a Two-species Fermionic System in the Presence of Feshbach Resonance }
\author{Raka Dasgupta}\email{rakadasg@bose.res.in}
\affiliation{S.N.Bose National Centre For Basic Sciences,Block-JD,
Sector-III, Salt Lake, Kolkata-700098, India}

\date{\today}
\begin{abstract}
We investigate the phenomenon of fermionic pairing with mismatched Fermi surfaces in a two-species system in the presence of Feshbach resonance, where the resonantly-paired fermions combine to form bosonic molecules. We observe that the Feshbach parameters control the critical temperature of the gapped BCS superfluid state, and also determine the range over which a  gapless breached 
pair state may exist. Demanding the positivity of the superfluid density, it is shown that although a breached pair state with two Fermi surfaces is always unstable, its single Fermi-surface counterpart can be stable if the chemical potentials of the two pairing species have opposite signs. This condition is satisfied only over a narrow region in the BEC side, characterized by an upper and a lower limit for the magnetic field. We estimate these limits for a mixture of two hyperfine states of $^6$Li using recent experimental data.

\end{abstract}


\maketitle
\section*{1. INTRODUCTION}

\indent The problem of fermionic pairing and superfluidity with mismatched Fermi surfaces has been widely investigated in recent years from theoretical standpoints \cite{liu,bed,wu1,cald,carl,son,sheehy,sheehy2,chien,chen,kop,iskin,gubb} as well as experimentally \cite{zw1,zw2,pat}.

A mismatch in Fermi surfaces can be easily realized in a two-species fermionic system where the pairing species have unequal populations or different masses/ chemical potentials. Typically, such a system would consist of two different fermionic atoms (e.g., $^6$Li and $^{40}  $K) or alternatively, two hyperfine states of the same atom (e.g., states  $|F =1/2, m_F = 1/2\rangle$ and  $|F = 1/2, m_F = -1/2\rangle$ of $^6$Li  atoms).

 Several phases have been proposed to describe the possible ground state of such a system, including the FFLO phase \cite{LO, FF, manna}(pairing with non-zero centre-of-mass momentum, where the order parameter shows a spatial variation), the gapless BP (breached pair) phase \cite{liu, bed, wu1} (pairing with zero centre-of mass momentum: where the order parameter is non zero but the excitation energy becomes zero), and the inhomogeneous phase-separated state \cite{sheehy, sheehy2}, where any two pure states coexist.

 In this note, our focus is on the homogeneous and gapless breached pair state, also known as the Sarma state. Sarma \cite{sarma}, in the early studies of superconductivity, predicted a spatially isotropic, homogenous  and uniform state with gapless excitation modes in the presence of a magnetic field. However, for weak coupling BCS theory, this gapless breached pair state marks the maximum of the thermodynamic potential, and thus, cannot be the stable ground state of the system. This is the well-known Sarma instability. In the last few years, several mechanisms were put forward to avoid this instability. According to Forbes et al. \cite{forbes1}, a stable Sarma state is possible in a model with finite range interaction where the momentum dependence of the pairing gap cures the instability. It has also been proposed by a number of workers \cite{ pao, son,  hu1, manna} that the breached pair state becomes stable in the deep BEC regime, if the BCS-BEC crossover picture is taken into account. He et al.,in a very recent work \cite{he3} has argued that the breached pair state can be a possible ground state in the weak coupling region for a two-band Fermi system.\\
 \indent To study the breached pair state,  we start with  a two-species fermionic system. In addition to the weak BCS attraction ( denoted by $-g_1$), we consider a strong interaction ($g_2$) of the Feshbach variety which couples  a fermion of type $a$ with a $b$ fermion to form a bosonic molecule $B$. Our model resembles the one used in \cite{ohashi, holland}, but we extend it to cover the two-species case. The system is described by the Hamiltonian:

\begin{equation}
\label{hamil}
 H=\sum(2\nu- \mu_B)B_0^{\dagger}B_0  +   \sum{\Tilde\epsilon_p^a a_p^{\dagger}a_p}  +  \sum{\Tilde\epsilon_p^b b_p^{\dagger}b_p} - g_1 \sum a_{p'}^{\dagger}b_{-p'}^{\dagger}b_{-p}a_p +  g_2 \sum[ B_0^{\dagger}a_{p}b_{-p} + {a^{\dagger}_{p}}b^{\dagger}_{-p}B_0 ]
 \end{equation}
 
 Here $a_p$, $a^{\dagger}_p$ are the creation and annihilation operators for atom $a$, while $b_p$, $b^{\dagger}_p$ are the corresponding operators for atom $b$.  Also, $ \tilde \epsilon_p^a$ = $\epsilon_p^a-\mu_a$ and $ \tilde \epsilon_p^b$ = $\epsilon_p^b-\mu_b$, while  $\epsilon^a_p$,  $\epsilon^b_p$ are the respective kinetic energies and $\mu_a$, $\mu_b$ the respective chemical potentials for species $a$ and $b$. The annihilation and creation operators for the composite boson $B$ are $B_0$ and $B^{\dagger}_0$ (We restrict ourselves to the case where only zero-momentum bosons are formed).The chemical potential for the bosons is $\mu_B$, and $2\nu$ is the threshold energy of the composite bose particle energy band.\\

Using this Hamiltonian, we study the effect of the Feshbach parameters first on the gapped BCS superfluid state, and then on the gapless Sarma state, taking a variational approach. The stability of such a gapless state is analysed ensuring the positivity of the superfluid density \cite{he1, he2}. Considering the BCS-BEC crossover picture, we show that the breached pair state is stable only in a narrow region in the BEC side, bounded by two magnetic field values.

\section*{2. GAP EQUATION AND GAPLESS EXCITATIONS}

We want to study the ground state of the two-fermion system by a variational method. We take $|\Psi\rangle=|F\rangle\otimes|B\rangle$
as the trial form of the ground state of the system. Here $|F\rangle$ =$|BCS\rangle$ =$\prod(U_p+V_p a^{\dagger}_pb^{\dagger}_{-p})|0\rangle$ , i.e, the probability of the pair ($a_{p\uparrow}$, $b_{-p\downarrow}$)being occupied is $|V_p|^2$, and the probability that it is unoccupied is $|U_p|^2= 1 -|V_p|^2$ , while $|B\rangle$ is the ground state for the condensate part of the boson subsystem. Our variational prescription is that $|B\rangle$ has to be chosen in such a manner that it is an eigenstate of the annihilation operator B, which would make it a coherent state. In terms of Fock states $|B\rangle$ =$\Sigma C_n|n\rangle$. Using the normalization condition, we arrive at $|B\rangle =\mbox{exp}(-\alpha^2/2)\mbox{exp}(\alpha B^{\dagger})|0\rangle $. The ground state of the system is consequently given by
\begin{equation}
\label{grnd}
|\Psi\rangle = \prod (U_p+V_p a^{\dagger}_pb^{\dagger}_{-p})|0\rangle\otimes \mbox{exp}(-\alpha^2/2+\alpha B^{\dagger})|0\rangle 
\end{equation} 
where $\alpha = \sqrt{N_B}$, $N_B$ being the expectation value of the total number of bosons in the condensed state.
 
\indent From the Hamiltonian (\ref{hamil}) and the ground state wave function (\ref{grnd}), the ground state energy of the system would be

 \begin{equation}
 E= \sum( \Tilde \epsilon_p^a+ \Tilde \epsilon_p^b) V_p^2 - g_1\sum_{p,p'} U_pV_pU_{p'}V_{p'} + (2\nu-\mu_B)\alpha^2 + 2g_2\alpha\sum_p U_pV_p
 \end{equation}
 
 Minimizing $E$ with respect to $V_p$ and $\alpha$ we get

\begin{equation}
\label{gap}
4\epsilon^{+}_p V_p -2g_{eff}\sum_{p,p'}U_{p'}V_{p'}(U_p-V_p^2/U_p) =0
\end{equation}
where $\epsilon^{+}_p =(\Tilde \epsilon_p^a+ \Tilde \epsilon_p^b)/2$. \\
$g_{eff}$ is defined by the relation
\begin{equation}
\label{geff}
g_{eff}=g_1+g_2^2/(2\nu-\mu_B)
\end{equation}

We note that this expression matches with the one obtained by Ohashi et al \cite{ohashi} using  diagrammatics. In order to avoid a possible ultraviolet divergence, the summation in the right hand side of equation(\ref{gap}) has to be carried over upto a cutoff. For a metallic superconductor, this cutoff is played by $\hbar\omega_D$, $\omega_D$ being the Debye frequency. In ultracold atomic systems, the cutoff is determined by the range of the interatomic potential.

Now, if we choose $g_{eff}\sum_{p'}U_{p'}V_{p'} =\Delta$, the usual form of the gap equation follows, and we can identify $\Delta$ as the gap in the excitation spectrum. A similar relation is obtained in finite temperature systems as well, if we incorporate the appropriate Fermi distribution functions in the expression of the free energy. In terms of this gap parameter, $U_p$ and $V_p$ can be expressed as $U_p^2=\frac{1}{2}\left(1+\frac{\epsilon_p^{+}}{\sqrt{{\epsilon_p^{+}}^2+\Delta^2}}\right)$ and $V_p^2=\frac{1}{2}\left(1-\frac{\epsilon_p^{+}}{\sqrt{{\epsilon_p^{+}}^2+\Delta^2}}\right)$.\\
\indent The critical temperature $T_C$ (temperature at which the gap vanishes) is proportional to $\Delta_0$, the BCS gap in the weak coupling limit, which is given by \cite{pethick}
\begin{equation}
\Delta_0 = \frac{8}{e^2}\epsilon_F  \mbox{exp}(-\frac{1}{\rho(0)g_{eff}})
\end{equation}
Here $\rho(0)$ is the density of states in the Fermi level, $e$ is the base of natural logarithms, and $\epsilon_F$ is the Fermi energy.  
Thus the value of $\Delta_0$ ( and $T_c$) can be raised or lowered by adjusting  $2\nu$, the tunable parameter depending on the Feshbach resonance process.\\
\indent Next we study gapless excitations.The quasiparticle dispersions as obtained from standard BCS-like treatment is of the form \cite{sed, wu1}:
\begin{equation}
E^{a,b}_p = \pm\frac{\Tilde\epsilon^a_p -\Tilde\epsilon^b_p}{2} +\sqrt{{(\frac{\Tilde\epsilon^a_p +\Tilde\epsilon^b_p}{2}})^2 +\Delta^2}
\end{equation}
\begin{figure}
\includegraphics[scale=.7]{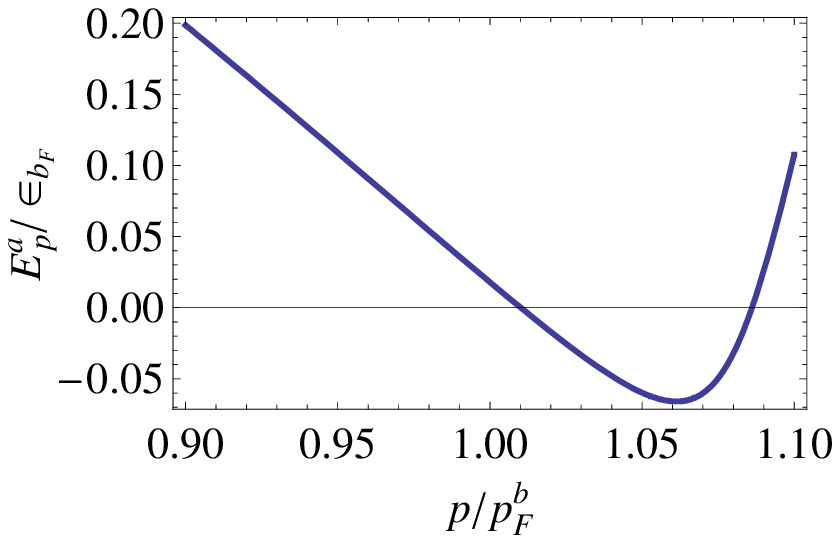}
\includegraphics[scale=.7]{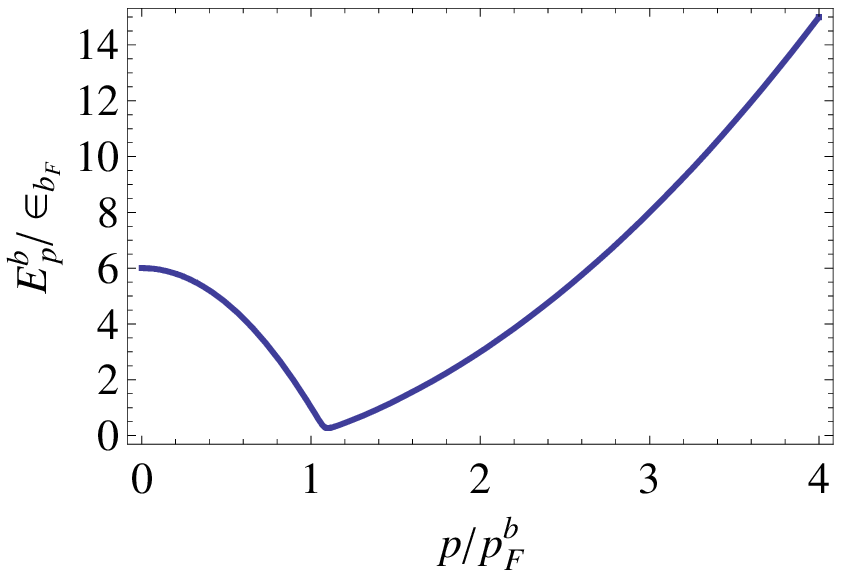}
\caption{($E_p^{a,b}$-$p$) curve when the Feshbach term is absent}
\includegraphics[scale=.7]{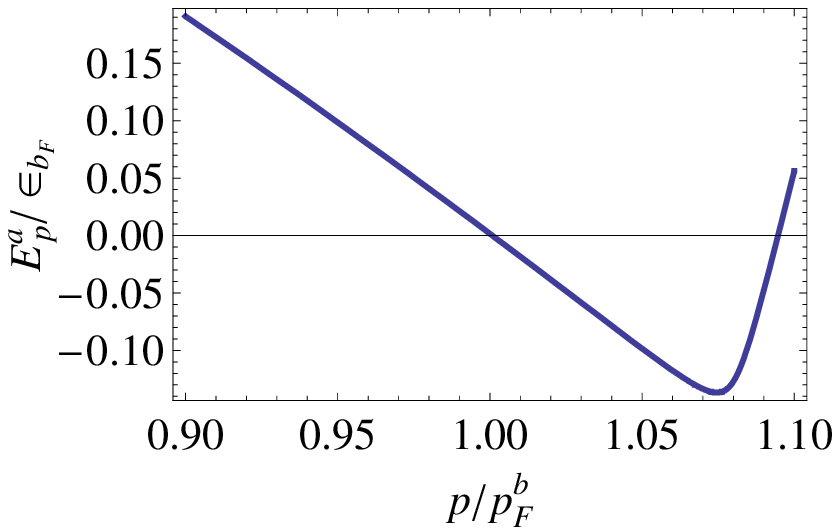}
\includegraphics[scale=.7]{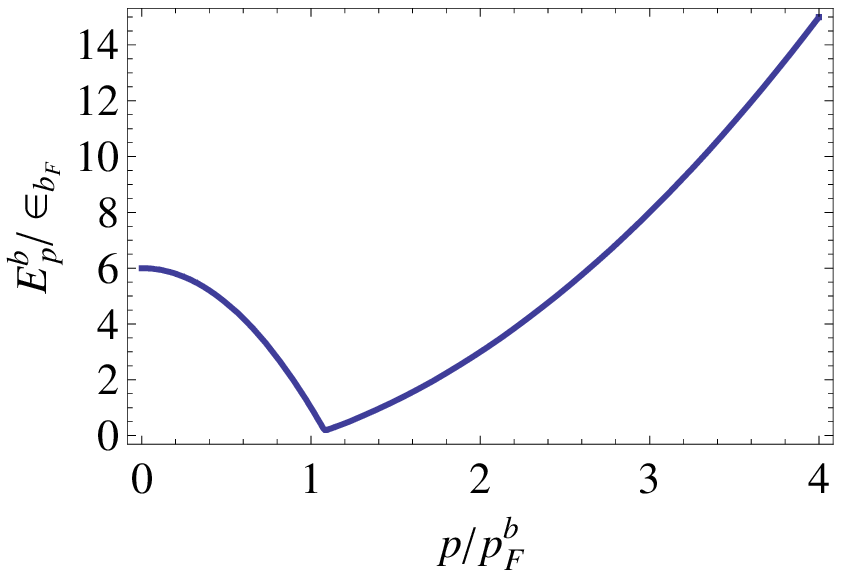}
\caption{($E_p^{a,b}$-$p$) curve for $\Tilde{\mu_B}-\Tilde{2\nu}=.5$, $\Tilde{g_2}=.2$}
\includegraphics[scale=.7]{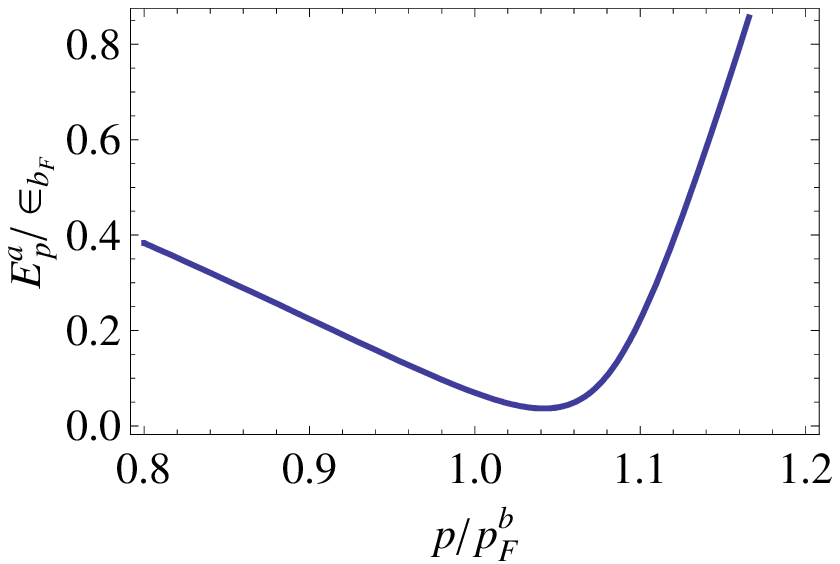}
\includegraphics[scale=.7]{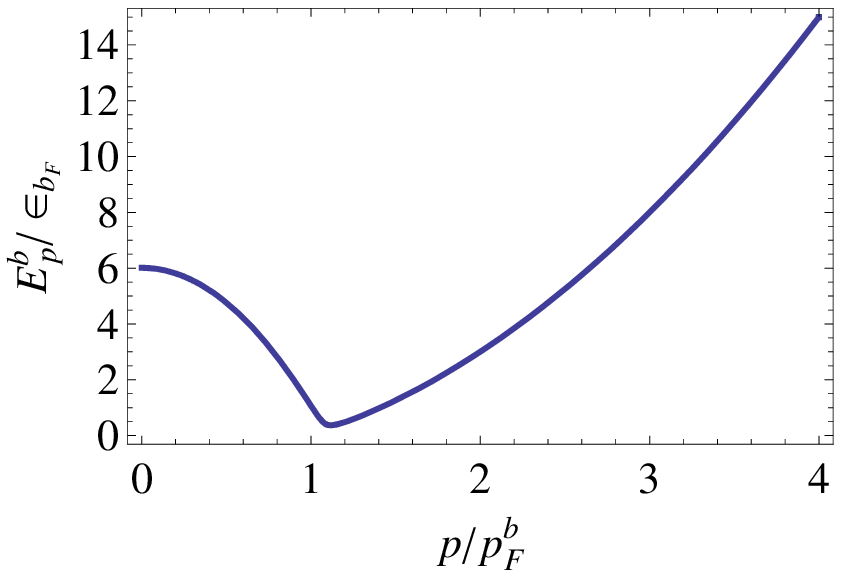}
\caption{($E_p^{a,b}$-$p$) curve for $\Tilde{\mu_B}-\Tilde{2\nu}=-.5$, $\Tilde{g_2}=.2$}
\end{figure}

Depending on the values of particle masses and corresponding chemical potentials, these quasiparticle excitations can be negative, thus leading to gapless excitations. This is possible only if the magnitude of $\Delta$ is less than a critical value $\Delta_c$ \cite{sed, wu1}.

\begin{equation}
\Delta_c =\frac{|m_b\mu_b -m_a\mu_a|}{2\sqrt{m_am_b}}
\end{equation}

When $|\Delta|>\Delta_c$, both $E^{a}_p$ and $E^{b}_p$ remain positive for all values of $p$. This corresponds to usual BCS pairing. When $|\Delta|<\Delta_c$, either $E^{a}_p$ or $E^{b}_p$ crosses zero at the points:

\begin{equation}
\label{p2-11}
p_{1,2}^2= (m_b\mu_b +m_a\mu_a)\mp [(m_b\mu_b -m_a\mu_a)^2-4m_am_b\Delta^2]^{1/2}
\end{equation}
The difference between $p_1$ and $p_2$ gives the span over which we get a gapless region in the parameter space. The state with gapless excitations marks the coexistence of the superfluid and normal components at zero temperature, and is called the Sarma phase, or the Breached Pair state. When we include the Feshbach term in the Hamiltonian, we have a control over this Sarma phase as well.

In figures(1,2,3), we plot the $E$-$p$ curve for a two-species system. Here we scale all energies by $\epsilon_{b_F}$ 
( Fermi energy of species $b$) and all momenta by $p_{b_F}$ (Fermi momentum of species $b$). It is evident that in this convention, $m_b=.5$ and $\mu_b=1$(in the BCS limit). We choose $m_a=.1$, $\mu_a=6$ and  multiply quantities $g_1$, $g_2$, $(2\nu-\mu_B)$ and $g_{eff}$ by $\rho(0)$, the density of states at the Fermi level to get dimensionless quantities $\Tilde{g_1}$, $\Tilde{g_2}$, $(\Tilde{2\nu}-\Tilde{\mu_B)}$ and $\Tilde{g_{eff}}$. Let $\Tilde{g_1}=0.3$.  Had there been no Feshbach coupling $g_2$, we would get a gapless region from $p$=1.01 to $p$=1.85 ( in units of $p_{b_F}$) as seen from Figure-1.\\
\indent If $\Tilde{g_2}=.2$ is introduced in the system, and we choose $\nu$ in such a way that $(\Tilde{\mu_B}-\Tilde{2\nu)}=0.5$, we get the Sarma phase for a wider region, from $p$=1.00 to $p$=1.95 (in units of $p_{b_F}$) as seen from Figure-2.\\
If, on the other hand, $(\Tilde{\mu_B}-\Tilde{2\nu)}=-.5$, the gapless phase vanishes entirely (Figure-3).\\

\section*{3. STABILITY ANALYSIS}

The stability of a superfluid phase has been studied in different ways like demanding positivity of the superfluid density \cite{pao},  ensuring non-negative eigenvalues of the number susceptibility matrix of the system\cite{pao, gubankova}, minimization of thermodynamic potential \cite{sheehy2, mishra} etc.

Here we adopt the first criterion and demand that the superfluid density $n_s$ must be positive. If the fermions are charged, this is equivalent to saying that the Meissner mass-squared  must be positive \cite{gubankova}, since $n_s$,  the superfluid density, and $M^2$, the Meissner mass squared obey the relation : $M^2=\frac{n_s q^2}{m^2}$ \cite{he1}, $q$ being the electronic charge and $m$, the mass of the particle. In our treatment, we work with the superfluid density directly, since we are talking of a charge neutral Fermi system. However, the term ``Meissner mass" can still be used just to continue the analogy with superconductors, as it only serves to express the superfluid density to within a multiplicative constant.

For convenience, we assume that the two species have equal masses. Therefore, equation (\ref{p2-11}) takes a simpler form:
\begin{equation}
\label{p2-12}
p_{1,2}={2m[\bar{\mu}\pm \sqrt{\delta\mu^2-\Delta^2}]}^{\frac{1}{2}}
\end{equation}
where $\bar{\mu}=(\mu_a+\mu_b)/2$ and $\delta\mu=(\mu_a-\mu_b)/2$.
Thus, $\Delta_c = \delta\mu$ here.

He et al \cite{he1,he2} have shown that when the two species have equal masses, the superfluid density can be expressed as 
\begin{equation}
\label{m2}
n_s=mn \left(1-\frac{\eta\delta\mu \theta(\delta\mu-\Delta)}{\sqrt{\delta\mu^2-\Delta^2}}\right)
\end{equation}

where $\eta=\dfrac{p_1^3+p_2^3}{6\pi^2 n}$. We note that here $n$ marks the bare fermion density or the density of the atoms that have not been part of the condensate yet  ( i.e, unpaired fermions, fermions forming the Cooper pair and fermions that constitute the non-condensate bosons all are  counted in n). Since the total number of bare Fermi atoms is conserved, so $n+ 2N_B$= constant, $N_B$ being the expectation value of the total number of bosons in the condensate. 
If $\Delta<\Delta_c$, for $n_s$ is to be positive, $\dfrac{\eta\delta\mu}{\sqrt{\delta\mu^2-\Delta^2}}$ has to be less than 1, which is satisfied if $\eta< 1$, and $|\eta^2-1|>\dfrac{\Delta^2}{\delta\mu^2}$. \\
Now when both $p_1$ and $p_2$ are real and there is a breached pair phase between them (this state has sometimes been termed as BP2 state in the literature \cite{yi, cald2}), the bare fermion density can be written as: 
\begin{equation}
\label{n1}
n =\dfrac{1}{2\pi^2}\displaystyle\int^{p_2}_{p_1}p^2\,dp+ \dfrac{1}{\pi^2}\left[\displaystyle\int^{p_1}_0V_p^2p^2\,dp+\displaystyle\int^\infty_{p_2}V_p^2p^2\,dp \right]
\end{equation}
Here the first term denotes the contribution from the gapless region between $p_1$ and $p_2$, i.e, the normal component. The next two integrals take care of the contributions from the gapped superfluid regimes, one from momenta $0$ to $p_1$ and the other from $p_2$ to infinity. The terms, when rearranged, gives
\begin{equation}
n= \frac{p_1^3+p_2^3}{6\pi^2}+ \frac{1}{\pi^2}(-\displaystyle\int^{p_1}_0 p^2 U_p^2\,dp+\displaystyle\int^\infty_{p_2} p^2 V_p^2\,dp)
\end{equation}
In the weak coupling limit, the last two integrals are very small, and hence, can be neglected. Therefore, $\eta$=1, and $n_s$ is negative. Thus, the Sarma state is unstable here.\\\\
We shift our focus to a special case, where only one of $p_1$ and $p_2$ is real ( in the literature, this is called the BP1 state \cite{yi,cald2}). In this case the $E$-$p$ curve resembles figure 4, and the bare fermion density is
\begin{equation}
n= \frac{p_2^3}{6\pi^2}+ \frac{1}{\pi^2}\displaystyle\int^\infty_{p_2} p^2 V_p^2\,dp
\end{equation}
This yields  $\eta<1$, and thus the stability criterion is fulfilled.
\begin{figure}
\includegraphics[scale=.7]{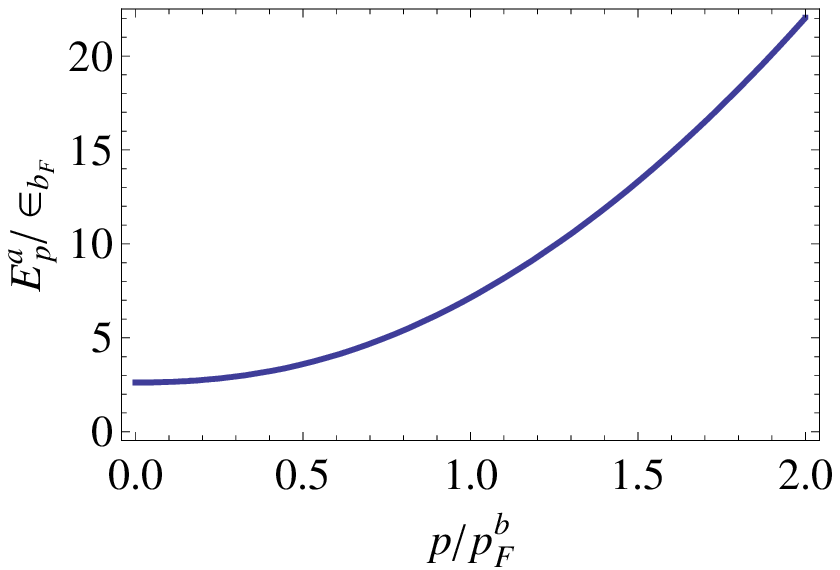}
\includegraphics[scale=.7]{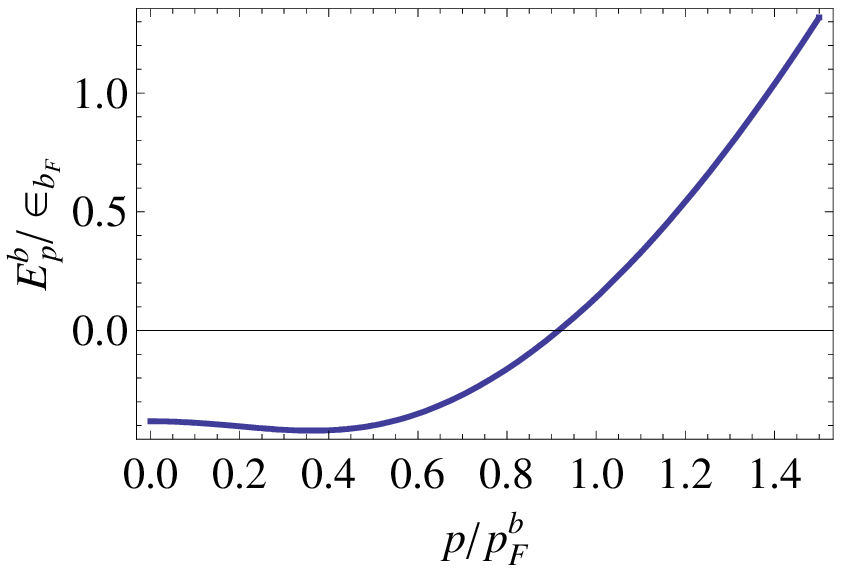}
\caption{($E_p^{a,b}$-$p$) curve when $p_1$ is imaginary }
\end{figure}
Now, from equation(\ref{p2-12}), $p_1$ is imaginary if $\sqrt{\delta\mu^2-\Delta^2}>\bar{\mu}$, i.e,  $-\mu_a\mu_b>\Delta^2$. So, Sarma phase is stable only in a region where the chemical potential of one species is positive, and the other, negative, provided the magnitude of their product is greater than 
$\Delta^2$.\\

Leggett has shown \cite{leggett} that in the BCS-BEC crossover picture, the chemical potential can be determined by solving the gap and the number equations. Extending these equations to the two species case, we get
\begin{subequations}
\begin{eqnarray}
\sum_{p=0}^{\infty}(\dfrac{1}{\epsilon_p}-\dfrac{1}{\sqrt{{\epsilon_p}^2+\Delta^2}})=\dfrac{m}{2\pi \hbar^2 a_s}\\
\sum_{p=p_2}^{\infty} (1-\dfrac{\epsilon_p-\bar{\mu}}{\sqrt{{\epsilon_p}^2+\Delta^2}})=2\dfrac{k_{b_F}^3}{3\pi^2}\\
\sum_{p=0}^{\infty} (1-\dfrac{\epsilon_p-\bar{\mu}}{\sqrt{{\epsilon_p}^2+\Delta^2}}) +\sum_{p=0}^{p_2} (1+\dfrac{\epsilon_p-\bar{\mu}}{\sqrt{{\epsilon_p}^2+\Delta^2}})=2\dfrac{k_{a_F}^3}{3\pi^2}
\end{eqnarray}
\end{subequations}

where $a_s$ is the $a-b$ scattering length and $k_{a_F}$ and $k_{b_F}$ correspond to the Fermi wave number of the more and the less populated species respectively.  Let $ k'$ denote the wave number corresponding to the breaching point for the BP1 state, i.e, $p_2 = \hbar k'$. Converting the sums into integrals, we obtain that in the weak coupling limit, $\bar{\mu}=(\epsilon_{a_F}+\epsilon_{b_F})/2=\hbar^2 (k_{a_F}^2+k_{b_F}^2)/4m$, as expected. In contrast, in the strong coupling limit, 
\begin{equation}
\bar{\mu}=-\dfrac{\hbar^2}{2ma_s^2}+\dfrac{2\epsilon_{b_F}(k_{b_F}a_s)}{3\pi}\left(1+\dfrac{2k'^3a_s^3}{\pi}\right)
\end{equation}
provided $k'^2<<\bar{\mu}$ (a condition which is satisfied if the population imbalance is small compared to the total population). In this case $\bar{\mu}$ asymptotically approaches $-\hbar^2/2ma_s^2$, i.e, half the binding energy of the molecule.
\\
If we think of $\mu_a$ and $\mu_b$ separately,  the first one differs from the other by the Fermi energy of the excess fermions. Moreover, in the presence of a magnetic field $H$, there is an asymmetry between the chemical potentials given by $m_BH$,  $m_B$ being the fermion magneton.
So we have 
\begin{equation}
\delta\mu = \dfrac{1}{2}\left(\dfrac{ \hbar^2 k'^2}{2m} +m_BH\right)
\end{equation}

\begin{figure}
\includegraphics[scale=.7]{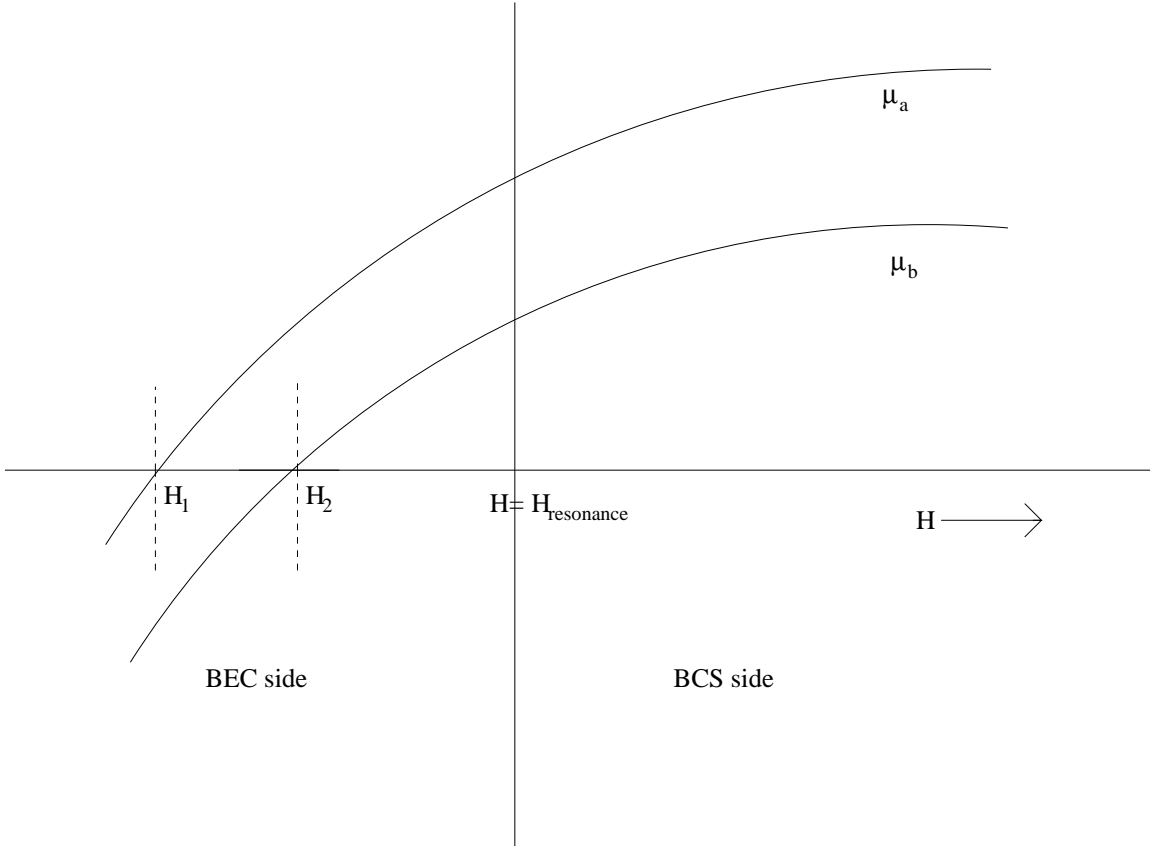}

\caption{Behavior of $\mu_a$ and $\mu_b$ in the crossover picture}
\end{figure}
Therefore, 
\begin{equation}
\label{mu12}
\mu_{a,b}=\bar{\mu}\pm\delta\mu = -\dfrac{\hbar^2}{2ma_s^2}+\dfrac{2\epsilon_{b_F}(k_{b_F}a_s)}{3\pi}\left(1+\dfrac{2k'^3a_s^3}{\pi}\right) \pm\left(\dfrac{ \hbar^2 k'^2}{4m} +\dfrac{m_BH}{2}\right)
\end{equation}

Our domain of interest is when $\mu_a$ is positive, and $\mu_b$ is negative, i.e, while approaching the BEC side, $\mu_b$ has already crossed zero but $\mu_a$ has not.
Now, $\mu_a$ and $\mu_b$ becomes zero at magnetic field values  $H_1$ and $H_2$ respectively, as sketched in figure (5), where
\begin{subequations}

\begin{eqnarray}
\label{h1}
H_1= (2/m_B)\left [ \frac{\hbar^2}{2ma_{s1}^2}-\frac{2\epsilon_{b_F}(k_{b_F}a_{s1})}{3\pi}\left(1+\frac{2k'^3a_{s1}^3}{\pi}\right) -\frac{ \hbar^2 k'^2}{4m}\right]\\
\label{h2}
H_2=(2/m_B)\left [ -\frac{\hbar^2}{2ma_{s2}^2}+\frac{2\epsilon_{b_F}(k_{b_F}a_{s2})}{3\pi}\left(1+\frac{2k'^3a_{s2}^3}{\pi}\right) -\frac{ \hbar^2 k'^2}{4m}\right]
\end{eqnarray}
\end{subequations}
Here $a_{s1}$, $a_{s2}$ are the respective values of the scattering length at $H_1$ and $H_2$. Between these two magnetic field values, the Sarma state will be stable. 

\section*{4. ESTIMATES OF $H_1$ AND $H_2$}

Experiments with population-imbalanced fermionic systems have been done by  Zwierlein et al. \cite{zw1, zw2} and Partridge et al \cite{pat}.They obtained the signature of superfluidity in an unequal mixture of two spin states of  $^6$Li atoms, and a quantum phase transition between the superfluid state and the normal state was observed at a critical polarization. Another method for experimental detection of the breached pair phase has been suggested by Yi et al. \cite{yi}. However, no clear signature of this gapless phase has been obtained till date.

We now use the data obtained from these experiments on population-imbalanced  gas of $^6$Li atoms  \cite{zw1, zw2} to make an estimate of the magnetic field values corresponding to the  breached pair Sarma state. To be able to use the expressions (\ref{h1},\ref{h2}), we need to know the scattering length as a function of the magnetic field. This is provided by $a_s =a_0 \left(1-\dfrac{\Gamma}{H-H_0}\right)$, where $a_0$ is the background scattering length, 
 $\Gamma$ is the width of the resonance, and $H_0$ is the position of the resonance peak. 
 
We use this expression for scattering length, and put $a_o=45.5r_0$ ($r_0$=Bohr radius), which is the singlet scattering length for $^6$Li. We also take $n_a= 1.8\times10^7$, $n_b=2.6\times10^6$ as population of the two species, $H_0$= 834 G,  $\Gamma$= 300G ( These values correspond to experimental data for Feshbach resonances in $^6$Li as reported in \cite{zw1,zw2,schunck}.) We now solve equations  (\ref{h1},\ref{h2}) explicitly to find $H_1$ to be 832.40 G and $H_2$ to be 833.95 G. It would be interesting to speculate that the superfluid observed by Zwierlein et al. in this range is of the BP1 variety.

This estimate of $H_1$ and $H_2$ is not a highly accurate one. In fact, the range of the breached pair state should get shifted a bit towards  lower values of the magnetic field. Actually, for deriving equations (\ref{h1},\ref{h2}),  we had assumed the magnitudes of the chemical potentials to be very large compared to $\Delta$ and $k_1^2$, an assumption which does not hold at points where $\mu_a$ and $\mu_b$ become zero. However, the values are not unreasonable in view of the fact that in the BCS-BEC crossover picture, the chemical potential falls quite sharply after crossing zero and quickly becomes a large negative quantity.

\section*{5. SUMMARY AND DISCUSSION}

Here we have studied a two-species fermionic system in the presence of Feshbach resonance, taking a variational route with an explicit construct of the ground state. The gapless breached pair state was discussed, and its stability was analysed ensuring the positivity of the superfluid density. We showed that a breached pair state with two Fermi surfaces is always unstable, while its single Fermi-surface counterpart is stable when the chemical potential of the two pairing species bear opposite  signs.

The stability of the breached pair state is indeed a widely debated issue. Although it has often been suggested that the BP1 state might be stable in deep BEC region, nothing, to the best of our knowledge, was said anything about how `deep' that really is. In this note, however, we observe that, the requirement that $\mu_{a}$ and $\mu_{b}$ should be of opposite signs, automatically puts two bounds in terms of the Feshbach magnetic fields, between which the gapless state is stable. Moreover, this stable breached pair state is obtained not in deep BEC, but near the vicinity of the point when the average chemical potential crosses zero, i.e, right after the onset of condensation.

Gubankova et al \cite{gubankova}, while discussing the stability of breached pair states by analysing the number susceptibility, commented that stable gapless states with a single Fermi surface exist for negative average chemical potential. In a recent paper, A. Mishra et al.\cite{mishra} reached the same conclusion by comparing the thermodynamic potentials of the condensed phase and the normal phase. Although the criterion they arrive at (the negativity of the average chemical potential) does not fully match with ours ( the chemical potential of the two species to have opposite signs), there is definitely a region of overlap.

In their experiment, Zwierlein et al. observed that superfluidity breaks down when the pairing gap $\Delta$ becomes small compared to the chemical potential difference $\mu_a-\mu_b$. We note that this matches with the stability criterion for the BP2 state, since, if $\Delta<\delta\mu$ and $\eta=1$, the state becomes unstable, as seen from equation (\ref{m2}).

As for the BP1 state,  we have shown that this state is stable in a region where   $\mu_a$ and $\mu_b$ have opposite signs, which can be achieved by keeping the system between two specific magnetic field values. For simplicity we took the two species to have same masses in our calculation, but the treatment should be extendable to a situation where the fermion species are of different masses, as for example in a $^6$Li-$^{40}$K system.
\\\\

\section*{ACKNOWLEDGEMENT}
The author would like to express her gratitude to her supervisor Prof. J. K. Bhattacharjee
for his keen interest and helpful comments during the course of this work. Constructive suggestions from the referee are also thankfully acknowledged.




\begin{thebibliography}{98}



\bibitem {liu} W. V. Liu and F. Wilczek, Phys. Rev. Lett. \textbf{90}, 047002 (2003).  
\bibitem {bed} P. F. Bedaque, H. Caldas, and G. Rupak, Phys. Rev. Lett. \textbf{91}, 247002 (2003).
\bibitem{wu1}S-T. Wu and S. Yip, Phys. Rev. A \textbf{67}, 053603 (2003)
\bibitem{cald} H. Caldas, C.W. Morais and A.L. Mota, Phys. Rev. D \textbf{72}, 045008 (2005)
\bibitem{carl} J. Carlson and S. Reddy, Phys. Rev. Lett. \textbf{95}, 060401 (2005).  
\bibitem{son} D. T. Son and M. A. Stephanov, Phys. Rev. A \textbf{74} 013614 (2006).
\bibitem{sheehy} D. E. Sheehy and L. Radzihovsky, Phys. Rev. Lett. \textbf{96}, 060401 (2006).
\bibitem{sheehy2} D. E. Sheehy and L. Radzihovsky, Annals of Physics \textbf{322}, 1790 (2007)
\bibitem{chien} C. C. Chien, Q. Chen, Y. He and K. Levin,  Phys Rev Lett. \textbf{98}, 110404, (2007)
\bibitem{chen} Q. Chen, Y. He, C. C. Chien and K. Levin,  Phys Rev B \textbf{75}, 014521, (2007)
\bibitem{kop}  T. Koponen, J. Kinnunen, J.P. Martikainen, L.M. Jensen, P. Torma,  New Journal of Physics B \textbf{8}, 179 (2006)
\bibitem{iskin} M. Iskin and C. A. R. Sa de Melo, Phys. Rev. Lett. \textbf{97}, 100404 (2006)
\bibitem{gubb} K. B. Gubbels, M. W. J. Romans, and H. T. C. Stoof,  Phys. Rev. Lett. \textbf{97}, 210402 (2006)
\bibitem{sed} Pairing in Fermionic Systems, edited by A. Sedrakian, J. W. Clark, and M. Alford, World Scientific Publishing Co, Singapore, 2006
\bibitem{sarma} G.Sarma, J.Phys.Chem.Solid 24,1029(1963)
\bibitem{LO} A. I. Larkin, Yu. N. Ovchinnikov, Sov. Phys.  JETP \textbf{20}, (1965)
\bibitem{FF} P. Fulde, R. A. Ferrel, Phys. Rev. A, \textbf{135}, 550, (1964)
\bibitem {forbes1} M. M. Forbes, E. Gubankova, W. V. Liu and F. Wilczek, Phys. Rev. Lett.\textbf{94}, 017001(2005)
\bibitem{hu1} H.Hu and X.J. Liu, Phys. Rev. A ,73, 051603(R)(2006)
\bibitem {manna}M. Mannarelli, G. Nardulli and M. Ruggieri, Phys. Rev. A.74, 033606(2006);
\bibitem{pao} C.-H. Pao, S.-T. Wu, and S.-K. Yip, Phys. Rev. B \textbf{73}, 132506 (2006).
\bibitem {mishra} A. Mishra and H. Mishra The Eur. Phys. J. D., \textbf{53},75,(2009)
\bibitem{he3} L. He and P. Zhuang , Phys. Rev. B \textbf{79}, 024511 (2009)
\bibitem{ohashi}Y. Ohashi and A. Griffin, Phys. Rev. Lett. \textbf{89}, 130402 (2002)
\bibitem{holland} M. Holland, S.J.J.M.F. Kokkelmans, M.L. Chiofalo, R. Walser, Phys. Rev. Lett.\textbf{87}, 120406 (2001)
\bibitem{feynman} R.P. Feynman, Statistical Mechanics, A Set of Lectures, Addison-Wesley, New York, 1988
\bibitem{pethick}C. J. Pethick and H. Smith, Bose-Einstein Condensation in Dilute Gases, Cambridge Univ. Press 2002
\bibitem{he1}L. He, M. Jin, P. Zhuang, Phys. Rev. B \textbf{73}, 214527 (2006)
\bibitem{he2}L. He, M. Jin, P. Zhuang, Phys. Rev. B \textbf{73},024511 (2006)
\bibitem{kita} M. Kitazawa, D.H. Rischke, I. Shovkovy, Phys. Lett. B \textbf {637}, 367 (2006)
\bibitem {gubankova} E. Gubankova, A. Schmitt, F. Wilczek, Phys. Rev. B \textbf{74}, 064505 (2006)
\bibitem{leggett}A.J. Leggett, Quantum Liquids,Bose Condensation and Cooper pairing in condensed-matter systems, Oxford University Press, Oxford, 2006
\bibitem{zw1} M. W. Zwierlein, A. Schirotzek, C. H. Schunck and W. Ketterle, Science \textbf{311}, 492 (2006)
\bibitem{zw2} M. W. Zwierlein, C.H. Schunck, A. Schirotzek, and W. Ketterle, Nature( London) \textbf{442}, 54 (2006)
\bibitem{pat}G.B. Partridge, W. Li, R. I. Kamar, Y. A. Liao, and R. G. Hulet, Science \textbf{311}, 503 (2006)
\bibitem{schunck} C. H. Schunck, M. W. Zwierlein, C. A. Stan, S. M. F. Raupach, and W. Ketterle, A. Simoni, E. Tiesinga, C. J. Williams, P. S. Julienne, PRA \textbf{71}, 045601(2005)
\bibitem{yi} W. Yi and L.-M. Duan, Phys. Rev. Lett. \textbf{97}, 120401 (2006)
\bibitem{cald2} H. Caldas and A.L. Mota, J Stat. Mech, P08013 (2008)






\end{thebibliography}
\end{document}